\documentstyle[eqsecnum,aps,pre,epsf]{revtex}
\newcommand{\bd}[1]{\mbox{\boldmath$#1$}}

\begin{document}

\twocolumn[\hsize\textwidth\columnwidth\hsize\csname@twocolumnfalse\endcsname

\title{
Strong Desynchronizing Effects of Weak Noise\\
in Globally Coupled Systems
}
\author{Jun-nosuke Teramae and Yoshiki Kuramoto}
\address{Department of Physics, Graduate School of Sciences, Kyoto University,
Kyoto 606-8502, Japan}
\maketitle

\widetext

\begin{abstract}
 In assemblies of globally coupled dynamical units, weak noise perturbing
 independently the individual units can cause anomalous 
 dispersion in the synchronized cloud of the units in the phase space. 
 When the noise-free  dynamics of the synchronized assembly is nonperiodic, 
 various moments of the linear dimension  of the cloud as a function 
 of the noise strength exhibit multiscaling  properties with 
 parameter-dependent scaling exponents. Some numerical evidence of this 
 peculiar behavior as well as its interpretation in terms of a 
 multiplicative stochastic process with small additive noise is provided.
 Universality of the phenomenon is also discussed.
\end{abstract}


]
\narrowtext

\section{Introduction}\label{sec.intro}

Oscillatory behavior, periodic or nonperiodic, observable on a macroscopic 
scale may often be a result of collective synchronization of a population 
of micro-oscillators\cite{winfree80,kuramoto84,heagy94b}. Consider, for
simplicity, a population of 
identical oscillators with global coupling. Systems of globally and
uniformally coupled 
oscillators in fact serve as a natural model for Josephson junction 
arrays\cite{hadley88,wiesenfeld96}, multimode lasers\cite{wang88,wiesenfeld90} and various biological systems such as 
flashing swarms of fireflies\cite{winfree67,buck81,buck88} and pacemaker cells responsible for 
circadian rhythms\cite{winfree67}. Under suitable conditions, the whole population
will exhibit complete synchrony when the oscillators are
free from intrinsic or extrinsic randomness. 
\par
In practical situations, however, 
the oscillators would be more or less noisy or non-identical, so that, instead of exhibiting 
perfect synchrony, they will be distributed in the phase space to form 
a cloud of a finite extension.
It seems quite natural to expect that, as far as the noise remains 
sufficiently weak, the long-time average of the linear dimension of this
cloud, denoted by  
$\langle r\rangle$, should be proportional to the noise strength.
This is actually the case when the oscillators are of the limit cycle type.
We shall find that, when the oscillators are chaotic, in contrast, such 
linear dependence breaks down under broad conditions, and is replaced by 
a more general power-law dependence. The main goal of the present paper 
is to explain theoretically why such peculiar behavior is possible for 
globally coupled chaotic oscillators. This will be verified with some 
numerical evidence for a number of population models. Our theory, which is 
quite similar to the one developed earlier to explain a certain unique feature 
of turbulence in nonlocally coupled
systems\cite{kuramoto90,kuramoto96,kuramoto97a,kuramoto97b,kuramoto98},
suggests that the general moments $\langle r^{q} \rangle$ as a function 
of the noise strength exhibits 
a simple multiscaling law. As a natural consequence from the same theory, 
similar multiscaling behavior is expected to occur even for the populations 
of limit cycle oscillators provided that, apart from the random noise,
the oscillators are subjected to another random force which is common
over the whole population. The crucial point here is that the average
motion has to be stochastic for the multiscaling behavior to occur. 
\par
In Sec.\ref{sec.model}, we start with a general class of 
differential-equation models for globally coupled elements with additive 
noise. Then, as a specific model belonging to this class, we study numerically 
a population of the chaotic R\"ossler oscillators\cite{rossler76}, and show the 
anomaly in the size of the synchronized cloud of the oscillators in the 
phase space. In Sec.\ref{sec.origin}, we develop a theory to explain the origin of a 
general scaling law including the results of Sec.\ref{sec.model} as a special case. 
Our theory becomes almost identical with the one developed in Refs.\cite{kuramoto96}
if the effects of the external noise are replaced by the effects of spatial
nonuniformity. The arguments in Sec.\ref{sec.origin} suggest that the phenomenon of 
concern should be so universal that a number of restrictions imposed on
our model could be removed. We provide some evidence for this in Sec.\ref{sec.summ}, 
where three different types of population model will be discussed. The final 
section summarizes our main conclusions.

\section{Model and numerical simulation}\label{sec.model}

Consider a population of $N$ identical units with the intrinsic dynamics 
given by $\dot{\bd{X}}=\mbox{\boldmath
$F$}\left(\mbox{\boldmath$X$}\right)$, where N is an arbitrary number. 
Typically, the dynamical units considered are chaotic oscillators. Introducing 
all-to-all type linear coupling and also external additive noise driving 
the oscillators individually, we have the following system of coupled 
differential equations:
\begin{eqnarray}
 \frac{d\mbox{\boldmath $X$}_i}{dt}
  &=& \mbox{\boldmath $F$}\left(\mbox{\boldmath $X$}_i\right)
  +K\cdot \left[\overline{\mbox{\boldmath $X$}}\left(t\right)-\mbox{\boldmath $X$}_i\left(t\right)\right]
  +f\cdot \mbox{\boldmath $\eta$}_i\left(t\right) \nonumber\\
  && \quad \left(i=1,\ldots ,N\right).
  \label{eq.dif}
\end{eqnarray}
Here $\bar{\bd{X}}$ is the simple average of $\bd{X}_i$ over the population, 
i.e., 
\begin{equation}
 \overline{\mbox{\boldmath $X$}}\left(t\right)
  \equiv\frac{1}{N}\sum_{i=1}^N{\mbox{\boldmath $X$}_i\left(t\right)}
  \label{eq.defXave},
\end{equation}
and $K$ is a positive coupling constant, thus working in favor of synchronous 
behavior of the population; $\bd{\eta}_{i}(t)$ represents the external 
noise with suitably normalized intensity applied independently on the 
oscillators, so that the coefficient $f$ measures the intensity of the noise.
For the sake of simplicity, the coupling term has been so arranged that the dynamics 
of a given unit which is in perfect synchrony with the average motion 
would be identical with its own dynamics without coupling. The coupling 
strength $K$ may be generalized to a coupling matrix, without qualitatively 
affecting the whole arguments which follow.
\par
In the absence of noise, the population will be attracted to a perfectly 
synchronized motion for sufficiently large $K$. In the one-oscillator 
phase space, the dynamics could then be imagined as an evolution of a single point.
When a small random noise is introduced, the resulting dynamics can no 
longer be represented by a point in the phase space. Instead, we have a 
cloud of a finite extension whose shape will be changing variously in time.
The effects of noise on the collective dynamics could conveniently be 
characterized  by the average linear dimension of the oscillator cloud in the 
phase space defined by 
\begin{equation}
 r\left(t\right)
  \equiv \frac{1}{N}\sum_{i=1}^N{\left|\overline{\mbox{\boldmath $X$}}\left(t\right)-\mbox{\boldmath $X$}_i\left(t\right)\right|}.
  \label{eq.rt}
\end{equation}
We define its various moments by a long-time average of 
$\langle r^{q}\rangle$:
\begin{equation}
 \left<r^{q}\right>\equiv\lim_{T\to\infty}\frac{1}{T}\int_0^Tr^{q}
 \left(t\right)dt~.
  \label{eq.defr}
\end{equation}
Our main concern below is the dependence of $\langle r^{q} \rangle$ on $f$, 
$K$ and $q$. Our intuition may tell that $\langle r^{q} \rangle$ will 
simply be proportional to $f^q$, which seemed to be true from some numerical 
simulation\cite{zanette98}. However, we see below that this is not always valid.
Instead, $\langle r^{q} \rangle$  behaves under broad conditions like 
\begin{equation}
\langle r^{q} \rangle\propto f^{\alpha(q)}
\label{eq.rq}
\end{equation}
for sufficiently small $f$, where the exponent $\alpha(q)$ is not linear 
in $q$ and also changes continuously with $K$ and other system parameters.
\par
\begin{figure}
\centerline{ \epsfxsize=9cm\epsfbox{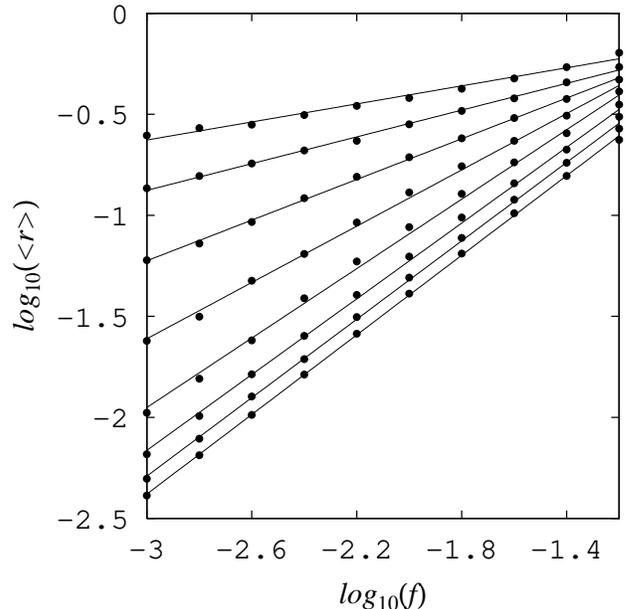}}
  \caption{
  $\log{\left<r\right>}$ vs. $\log{f}$ calculated numerically from
  Eq. (\ref{eq.ros}) with $N=64$ for several values of
  $K$. The top line corresponds to $K=0.142$ and the bottom one to
  $K=0.170$, with the uniform interval of 0.004.}
  \label{fig.ros.size}
\end{figure}
\begin{figure}
\centerline{ \epsfxsize=9cm\epsfbox{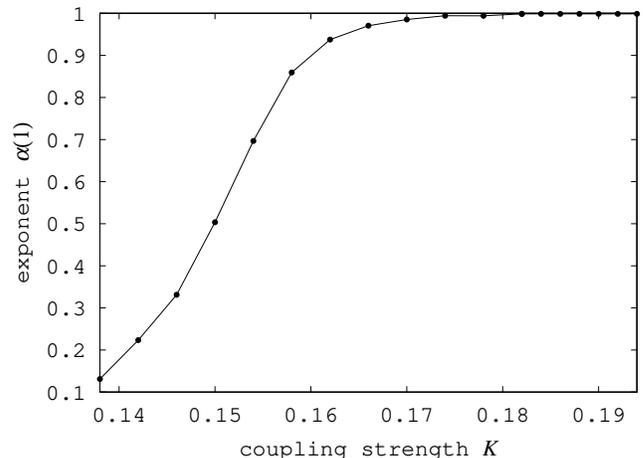}}
  \caption{Exponent $\alpha\left(1\right)$ vs. coupling strength $K$,
  obtained from the data of Fig. \ref{fig.ros.size} using the method of least-squares.}
  \label{fig.ros.pow}
\end{figure}

The above power-law dependence can be confirmed by a numerical analysis of 
globally coupled R\"ossler oscillators in the chaotic regime. By assuming
the coupling matrix to be diagonal, the system is governed  by the equations
\begin{equation}
 \left\{
  \begin{array}{l}
   \dot{X_i}=-\left(Y_i+Z_i\right)+K\cdot \left(\overline{X}-X_i\right)
   +f\cdot \eta_{i,X},\\
   \dot{Y_i}=X_i+aY_i+K\cdot \left(\overline{Y}-Y_i\right)
   +f\cdot \eta_{i,Y},\\
   \dot{Z_i}=b-cZ_i+X_iZ_i+K\cdot \left(\overline{Z}-Z_i\right)
   +f\cdot \eta_{i,Z},
  \end{array} 
	 \right.
	 \label{eq.ros}
\end{equation}
where the noise components $\eta_{i,\nu}$ ($\nu=X,Y,Z$) are assumed 
to be mutually independent and white Gaussian with vanishing mean and
unit variance. We will choose  
the standard parameter values $a=0.3$, $b=0.2$ and $c=5.7$. It turns out that 
in the absence of noise the state of perfect synchrony is stable above 
some critical value $K_c$ of $K$, where $K_c\simeq 0.4$. We will work 
with this condition throughout, and concentrate on the the size of the 
oscillator cloud when the noise sets in. Some numerical results for the first 
moment $\left<r\right>$ are summarized in Figs. \ref{fig.ros.size} and \ref{fig.ros.pow}.
In Figure \ref{fig.ros.size},  
log-log plots of $\left<r\right>$ versus $f$ are displayed for different 
values of $K$. The power-law behavior as given by Eq. (\ref{eq.rq}) with $q=1$ seems 
to hold well with the exponent $\alpha(1)$ changing with
$K$. Fig. \ref{fig.ros.pow} shows 
the dependence of the estimated values of $\alpha(1)$ on $K$. For
sufficiently
strong coupling, the exponent saturates to the normal value $\alpha(1)=1$, 
while for weaker coupling $\alpha(1)$ varies between 0 and 1. Since the noise 
is sufficiently small, the value of $\alpha(1)$ less than 1 implies anomalous 
amplification in the size of the oscillator cloud as compared with the normal 
case. 

\section{Origin of anomalous fluctuations}\label{sec.origin}

In this section, we present a theory on the origin of anomalous size 
fluctuations of a synchronized cluster found numerically in the preceding 
section. Our theory can be developed quite in parallel with the theory 
developed before on the multiscaled turbulence in nonlocally coupled 
systems\cite{kuramoto90,kuramoto96,kuramoto97a,kuramoto97b,kuramoto98}.
Let us start with Eq. (\ref{eq.dif}). Note that the average motion of the
population obeys the equation
\begin{equation}
 \frac{d\overline{\mbox{\boldmath $X$}}}{dt}
  =\mbox{\boldmath $F$}\left(\overline{\mbox{\boldmath $X$}}\right)
  +O\left(\frac{1}{N}\sum_{i=1}^N\left|\mbox{\boldmath $X$}_i
  -\overline{\mbox{\boldmath $X$}}\right|^2\right).
  \label{eq.ave}
\end{equation}
Thus, the deviation defined by $\mbox{\boldmath $x$}_i\left(t\right)
  \equiv \mbox{\boldmath $X$}_i\left(t\right)-
  \overline{\mbox{\boldmath $X$}}\left(t\right)$ 
is governed by
\begin{equation}
\frac{d\bd{x}_i}{dt}=\left(\bd{DF}\left(\bar{\bd{X}}\left(t\right)\right)-\bd{K}\right)\bd{x}_{i}
+f\cdot\eta_{i}(t)+O(|\bd{x}_{i}|^2),
\label{eq.dev}
\end{equation}
where $\mbox{\boldmath $DF$}\left(\overline{\mbox{\boldmath $X$}}(t)\right)$ 
is the Jacobian of $\mbox{\boldmath $F$} (\bd{X})$
at $\bd{X}=\overline{\mbox{\boldmath $X$}}\left(t\right)$. 
For sufficiently large $t$, the vector $\mbox{\boldmath
$x$}_i\left(t\right)$ is expected to become parallel with the Lyapunov
eigenvector of $\bd{DF}(\bar{\bd{X}}(t))-\bd{K}$ 
corresponding to the largest Lyapunov eigenvalue. By denoting the
amplitude of this 
eigencomponent of $\mbox{\boldmath $x$}_i\left(t\right)$ as 
$x_i\left(t\right)$, and dropping the suffix $i$ for the sake of simplicity,
Eq. (\ref{eq.dev}) is reduced to a scalar equation
\begin{equation}
 \frac{dx}{dt}
  =\lambda\left(t\right)x
  +f\cdot \eta\left(t\right)
  +O\left(x^2\right)~,
  \label{eq.xi}
\end{equation}
where $\lambda\left(t\right)$ is the local Lyapunov
exponent, and $\eta_{i}(t)$ is the projection of $\bd{\eta}_{i}(t)$ onto the
corresponding Lyapunov eigenvector. For the coupled R\"ossler oscillators 
given by Eq. (\ref{eq.ros}), we have
\begin{equation}
\lambda(t)=\lambda_{r}(t)-K,
\label{eq.lam}
\end{equation}
where $\lambda_{r}(t)$ is the local Lyapunov exponent of the individual 
R\"ossler oscillator. Since the sign of the long-time average of $\lambda(t)$, 
which will be denoted by $\lambda_{0}$, determines the stability of the 
perfect synchrony of the population in the absence of noise, the critical 
coupling strength for the coupled R\"ossler oscillators is simply given by
$K_{c}=\lambda_{0}$.
\par
Since we are working with the condition that the long-time average of 
$\lambda(t)$ is negative, and also with sufficiently weak noise, there should 
exits a range of $x$ where the nonlinear term of $O(x^2)$ as well as the 
noise term $f\cdot\eta(t)$ are negligible. Let this range be specified by 
$x_{min}\ll x\ll x _{max}$, where $x_{min}$ should be proportional to $f$.
In this range, we have a linear multiplicative stochastic equation
\begin{equation}
\frac{dx}{dt}=\lambda(t)x.
\label{eq.lamx}
\end{equation}
Note that the stochasticity here is of a deterministic origin. The stationary 
distribution from Eq. (\ref{eq.lamx}) can be discussed analytically in the following way.
Introducing a new variable $y$ by
\begin{equation}
y=\ln |x|,
\end{equation}
we rewrite Eq. (\ref{eq.lamx}) as 
\begin{equation}
\frac{dy}{dt}=\lambda(t).
\end{equation}
This is integrated to give
\begin{equation}
 y\left(t\right)-y\left(0\right)
  =\int_0^t\lambda\left(t\right)dt
  \equiv \Lambda\left(t\right)~.
  \label{eq.LAMBDA}
\end{equation}
The time-dependent probability distribution function (PDF) of $y$, denoted by 
$Q\left(y,t\right)$, satisfies
\begin{equation}
 Q\left(y,t\right)
  =\int_{-\infty}^{\infty} 
  \omega\left(\Lambda,t\right)Q\left(y-\Lambda,0\right)d\Lambda~,
  \label{eq.Q}
\end{equation}
where $\omega\left(\Lambda,t\right)$ is the PDF of $\Lambda\left(t\right)$ 
and is equivalent to the transition probability between the states at times 
0 and $t$. From the definition of $\Lambda$, $\omega\left(\Lambda,t\right)$ 
is expected to converge to a Gaussian of mean $t(\lambda_{0}-K)$ 
$(=t(K_{c}-K))$ and variance $tD_\lambda$ as $t\to\infty$, where
\begin{equation}
D_{\lambda}=\lim_{T\rightarrow\infty}\frac{1}{T}\left<\left\{\int_{0}^{T}
(\lambda(t)-\lambda_{0})dt\right\}^{2}\right>.
\end{equation}
Thus, the stationary PDF of $y$ becomes 
\begin{equation}
 Q\left(y\right)\propto e^{-\beta t}~,
  \label{eq.Qpro}
\end{equation}
where
\begin{equation}
 \beta= \frac{2\lambda_{0}}{D_{\lambda}}\,\,\,\,\,\left(>0\right)~.
  \label{eq.beta}
\end{equation}
Note that for the coupled R\"ossler oscillators, $\beta$ is given by
\begin{equation}
\beta=\frac{2(K-K_c)}{D_{\lambda}}.
 \label{eq.betaKc}
\end{equation}
The stationary PDF for the original variable $x$, denoted by $P(x)$, must
satisfy the relation $2P(x)dx=Q(y)dy$. Thus, from Eq. (\ref{eq.Qpro}), we obtain
\begin{equation}
 P\left(x\right)\propto x^{-\left(\beta+1\right)}
  \quad \left(x_{min}\ll x\ll x_{max}\right)~.
  \label{eq.P1}
\end{equation}
Coming back to the original stochastic equation (\ref{eq.xi}), the effects of additive 
noise and nonlinearity have now to be incorporated. There exists a 
characteristic value of $x$ below which the additive noise term
dominates the other terms. This value is what we denoted by $x_{min}$.
For $x\ll x_{min}$, the power-law divergence of $P(x)$ will be saturated to a 
constant. On the other hand, there exists the second characteristic 
value of $x$, denoted by $x_{max}$, above which the nonlinear term is 
the most dominant. If the nature of the nonlinearity is such that it 
decelerates rather than accelerates the growth of $x$, which we assume, 
the power-law decay of $P(x)$ will be replaced by a much sharper decay 
above $x_{max}$. As far as their dependence on $f$ is concerned,
$x_{min}$ and $x_{max}$ may be specified as
\begin{eqnarray}
x_{min}&=&f,\label{eq.xmin}\\
x_{max}&=&1 \label{eq.xmax}.
\end{eqnarray}
Thus, the entire profile of $P(x)$ will be such that it is nearly constant 
below $x=x_{min}$, obeys the above-mentioned power law, and decays quickly 
above $x=x_{max}$. On further idealization, we represent $P(x)$ by the 
analytic form
\begin{eqnarray}
 P\left(x\right)=\left\{
   \begin{array}{lcl}
    C{f}^{-(\beta+1)}& &\left(0<x<f\right) \\
    Cx^{-(\beta+1)}& &\left(f<x<1\right)\\
    0&,&\left(1<x\right) \\
   \end{array}
        \right.
	\label{eq.P}
\end{eqnarray}
where $C$ is a normalization constant. 
\begin{figure}
\centerline{ \epsfxsize=9cm\epsfbox{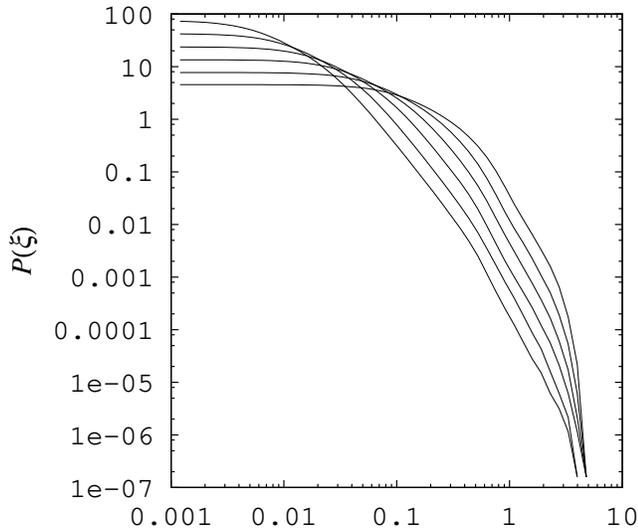}}
  \caption{
  Stationary PDFs of $P\left(\xi\right)$
  calculated numerically from Eq. (\ref{eq.ros}) for several values of
  noise strength $f$, where $K=0.16$ is fixed. The curves correspond to
  $f=10^{-1}$(lowest level at lower $\xi$), $10^{-1.25},
  10^{-1.5},10^{-1.75},10^{-2.00}$ and $10^{-2.25}$(highest).}
  \label{fig.ros.sta_f}
\end{figure}
\begin{figure}
\centerline{ \epsfxsize=9cm\epsfbox{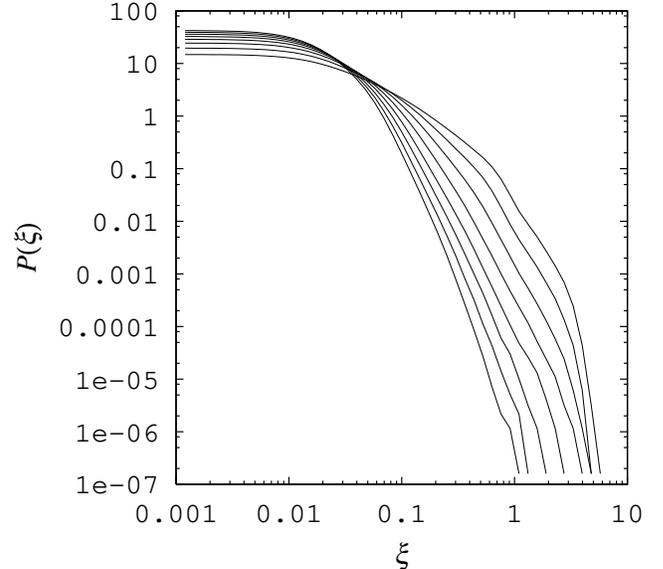}}
 \caption{Stationary PDFs $P\left(\xi\right)$ calculated numerically from
 Eq. (\ref{eq.ros}) for several values of coupling strength
 $K$, where $f=0.01$ is fixed. The curves correspond to
 $K=0.150$(slowest decay), 0.155, 0.160, 0.165, 0.170, 0.175 and
 0.185(fastest decay).}
  \label{fig.ros.sta_k}
\end{figure}
\par
Figures \ref{fig.ros.sta_f} and \ref{fig.ros.sta_k} show log-log plots
of $P(\xi)$ for the 
deviation $\xi$ of the first component $X$ obtained numerically from 
the coupled R\"ossler oscillators (Eq. (\ref{eq.ros})) for some values of $f$ and $K$.
It is seen that each curve is approximately composed of three parts 
corresponding to the three characteristic regimes in Eq. (\ref{eq.P}).
The range of $\xi$ where $P(\xi)$ is nearly constant extends proportionally
to $f$, which is consistent with Eq. (\ref{eq.P}), while the sharp drop of 
$P(\xi)$ seems to occur at some $\xi$ independent of $f$, which is again
consistent with Eq. (\ref{eq.P}) with $x$ replaced by $\xi$.
\par
Up to this point, we have discussed the statistics of the deviation $x$ of a
single element from the mean motion of the population. In what follows, 
we will identify the statistics of $|x|$ with that of $r$, i.e., the 
deviation averaged over the whole population. As far as some qualitative 
features such as the power-law dependence of $\langle r\rangle$ on $f$ 
is concerned, this assumption seems to be justified because the dynamical 
units are driven by a common multiplier $\lambda(t)$ by virtue of the global 
nature of the coupling. 
\par
By using Eq. (\ref{eq.P}), it is thus straightforward to calculate various moments
of $r$. As for the first moment, we have
\begin{eqnarray}
 \left<r\right>
  \simeq\int_{-\infty}^{\infty}\left|x\right|P\left(x\right)dx
  \propto\left\{
	  \begin{array}{lcl}
	   f^{\beta}& &\left(0<\beta<1\right) \\
	   f^1& &\left(\beta>1\right) \\
	  \end{array}
     \right.~.
     \label{eq.r}
\end{eqnarray}
Comparing the above equation with the expression in Eq. (\ref{eq.betaKc}) for the 
coupled R\"ossler oscillators, one may now understand the reason for the 
observed anomalous power-law dependence of $\langle r \rangle$ on the external 
noise when the system is not too far from the critical point $K=K_{c}$.
\par
The observed change in the scaling exponent with $K$ shown in
Fig. \ref{fig.ros.pow} still 
deviates considerably from Eq. (\ref{eq.r}) with $\beta$ given by Eq. (\ref{eq.betaKc}). 
Specifically, the numerical results do not exhibit sharp changes near 
$\beta=0$ and $\beta=1$. Such discrepancy seems to be due to the fact 
that the range of validity of Eq. (\ref{eq.r}) in terms of $f$ shrinks to zero 
as $\beta$ approaches 0 or 1. A little more careful analysis shows that 
under fixed $f$, we have $\langle r\rangle\sim 1/|\ln f|$ as 
$\beta\rightarrow 0$, and $\langle r\rangle\sim |f^{\beta}\ln{f}|$ as 
$\beta\rightarrow 1$. 
\par
Similar calculations for the higher moments $\langle r^{q}\rangle$ are
also straightforward, and the results are simply given by
\begin{eqnarray}
 \left< r^{q}\right>\propto\left\{
	  \begin{array}{lcl}
	   f^{\beta}&\ &\left(0<\beta<q\right) \\
	   f^{q}&\ &\left(\beta>q\right) ~,\\
	  \end{array}
	  \right.
          \label{eq.Mn}
\end{eqnarray}
except for the weak logarithmic singularity mentioned above.
Thus, the anomalous fluctuations could be visible through higher moments
in the range of stronger coupling where no anomaly is visible through
lower moments. Specifically, for the coupled R\"ossler oscillators,
the range of $K$ where the $q$-th moment behaves anomalously is given by
\begin{equation}
 K_c<K<K_c+\frac{qD_{\lambda}}{2}~.
  \label{eq.km}
\end{equation}
\par
Near the critical point at which the average Lyapunov exponent vanishes,
Eq. (\ref{eq.xi}) is of the same form as the equation employed for discussing the
so-called {\em on-off intermittency}\cite{fujisaka85,fujisaka86,platt93,heagy94a} with noise\cite{platt94,hammer94}. Since the dynamics of 
$r(t)$ would qualitatively be the same as that of $x(t)$ of a representative 
oscillator,
this implies that noisy on-off intermittency could also be observed
in $r(t)$. This was confirmed numerically for the coupled R\"ossler oscillators,
though we will not report its details. The only thing to be remarked is that
the origins of the power law in the on-off intermittency and that of our 
present concern are completely independent.

\section{Universality of anomalous fluctuations}\label{sec.uni}

Up to the preceding section, our discussion has been based on the following
three assumptions:
\begin{itemize}
\item [1.] The system is described by a set of ordinary differential equations
\item [2.] The source of randomness working against complete synchrony
is represented by external additive noise 
\item [3.] The constituents of the population are intrinsically chaotic
\end{itemize} 
In what follows, we will show that neither of these assumptions is necessary, 
which implies that the anomalous behavior of concern would be quite general.

\subsection{Case of globally coupled maps}\label{subsec.map}

\begin{figure}
\centerline{ \epsfxsize=9cm\epsfbox{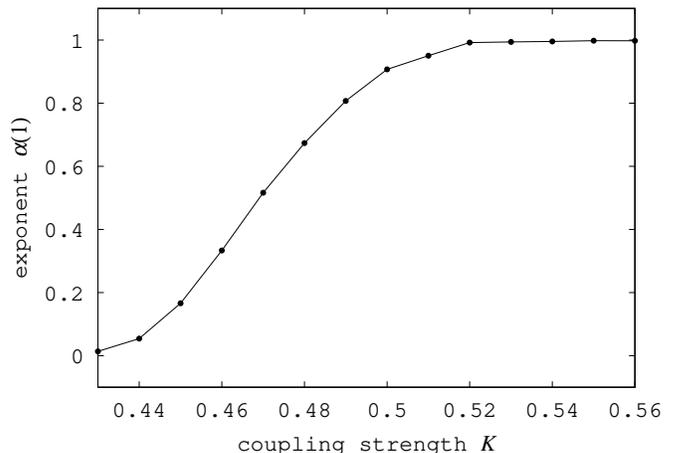}}
 \caption{Exponent $\alpha\left(1\right)$ vs. coupling strength $K$
 calculated numerically from the generalized tent map
 (Eq. (\ref{eq.gentent})) 
 with $a=0.75$}
  \label{fig.tent.pow}
\end{figure}

Discrete-time analogue to Eq. (\ref{eq.dif}) is the system of globally coupled 
maps\cite{kaneko89,kaneko90} with noise. Assuming for the sake of simplicity that the 
individual map is one-dimensional, we are concerned with the model equation
of the form
\begin{eqnarray}
 X_i\left(n+1\right)
  &=& \left(1-K\right) M\bigl(X_i\left(n\right)\bigr)
  +K M\left(\overline{X\left(n\right)}\right)
  +f\eta_i \nonumber\\
  && \quad \left(i=1,\ldots ,N\right).
  \label{eq.map}
\end{eqnarray}
One may develop arguments similar to the case of continuous-time dynamics, 
and derive a discretized version of Eq. (\ref{eq.xi}):
\begin{equation}
x_{n+1}=e^{\lambda_{n}}x_{n}+f\cdot\eta_{n}+O(x_{n}^2).
\label{eq.devmap}
\end{equation}
Furthermore, the arguments leading to a stationary PDF similar to the
form of Eq. (\ref{eq.P}) 
are almost the same as before. As an example, let us consider a generalized 
tent map for $M(x)$ defined by
\begin{eqnarray}
 M\left(x\right)=\left\{
   \begin{array}{lcl}
    x/a& &\left(0\leq x\leq a\right) \\
    \left(1-x\right)/\left(1-a\right)& &\left(a\leq x\leq 1\right)~. \\
   \end{array}
        \right.
	\label{eq.gentent}
\end{eqnarray}
The crucial condition for the occurrence of anomaly of $\langle r\rangle$ 
is that the local Lyapunov exponent in Eq. (\ref{eq.devmap}) fluctuates between positive 
and negative values. Thus, the conventional tent map ($a=0.5$) for which 
$\lambda=\ln2$ identically is ruled out.  
In a suitable range of $a$ and $K$, the system given by Eq. (\ref{eq.map}) is 
confirmed to exhibit power-law fluctuations in the form of Eq. (\ref{eq.r}).
Estimated exponent $\alpha(1)$ of the first moment of $r$ as a function of
$K$ is displayed in Fig. \ref{fig.tent.pow}. 

\subsection{Effects of heterogeneity}\label{subsec.inho}

Any kind of heterogeneity present in the population could give a source 
of incompleteness in synchrony. We focus on the following inhomogeneous 
system of elements {\em without} additive noise.
\begin{equation}
 \frac{d\mbox{\boldmath $X$}_i}{dt}
  =\mbox{\boldmath $F$}
  \left(\mbox{\boldmath $X$}_i,a_i\right)
  +K\cdot \left(\overline{\mbox{\boldmath $X$}}\left(t\right)
	   -\mbox{\boldmath $X$}_i\left(t\right)\right)~.
	   \label{eq.para}
\end{equation}
Here the heterogeneity is represented by a weak distribution in a parameter 
$a$, its value for the i-th element being denoted by $a_i$. The dynamics 
of the deviation $\bd{x}$ of a representative oscillator is described by
\begin{eqnarray}
 \frac{d\mbox{\boldmath $x$}}{dt}
  &=& \Bigl(\mbox{\boldmath $DF$}
    \left(\overline{\mbox{\boldmath $X$}}\left(t\right)
     ,\overline{a}\right)
    -\mbox{\boldmath $K$}\Bigr)\mbox{\boldmath $x$} \nonumber\\
    &+& \delta a\cdot \frac{\partial \mbox{\boldmath $F$}
    \left(\overline{\mbox{\boldmath $X$}}\left(t\right),\overline{a}\right)}{\partial a}
  +O\left(\left|\mbox{\boldmath $x$}\right|^2,\delta a^2\right)~,
  \label{eq.xi_para}
\end{eqnarray}
where suffix $i$ is dropped, and $\delta{a}$ denotes the deviation of $a$
from its population average $\bar{a}$. The effect of the second term
on the right-hand side is essentially the same as that of the additive noise
in  Eq. (\ref{eq.dev}), thus resulting in the anomalous power-law dependence of 
$\langle r^{q}\rangle$ on the strength of inhomogeneity. 

\subsection{Populations of non-chaotic units}\label{subsec.nonchao}

\begin{figure}
\centerline{ \epsfxsize=9cm\epsfbox{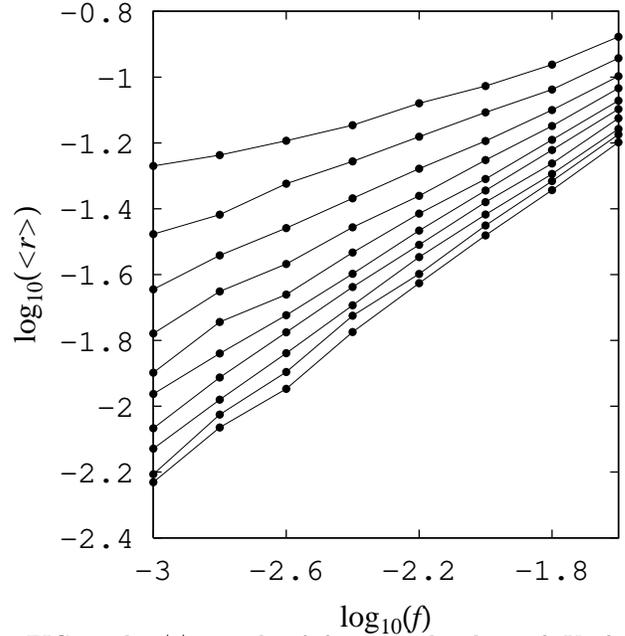}}
 \caption{$\log{\left<r\right>}$ vs. $\log{f}$ for several values of
 $K$, from $0.01\left(\mbox{top}\right)$ to
 $0.19\left(\mbox{bottom}\right)$ with the uniform increment of 0.002,
 calculated numerically from
 Eq. (\ref{eq.FN}) and (\ref{eq.ranwalk}); $N=64$,
 $\left(a,b,\epsilon\right)=\left(1.0,0.58,0.1\right)$, 
 $\left(G_0,\gamma,g\right)=\left(0.5,1.0,5.0\right)$.}
 \label{fig.fn.size}
\end{figure}

\begin{figure}
\centerline{ \epsfxsize=9cm\epsfbox{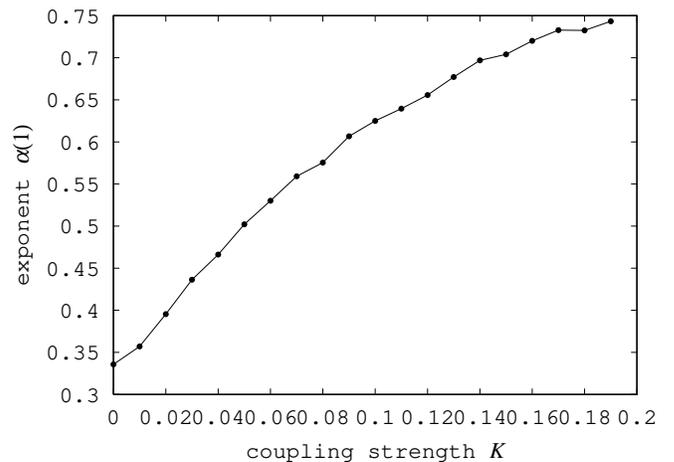}}
  \caption{Exponent $\alpha\left(1\right)$ vs. coupling strength $K$
  obtained from the data of Fig.(\ref{fig.fn.size}) using the method of least-squares.}
  \label{fig.fn.pow}
\end{figure}

We have seen that the anomalous fluctuations exhibited by the synchronized 
cluster can arise only if the local Lyapunov exponent associated with the 
individual unit fluctuates randomly. This means that we can never expect such 
anomaly for populations of {\em non-chaotic} dynamical units. Even for this 
class of systems, however, the anomalous fluctuations may arise provided 
the whole population is driven externally by a common random force apart 
from the random noise considered previously. The system of this class will 
take the form
\begin{equation}
 \frac{d\mbox{\boldmath $X$}_i}{dt}
  =\mbox{\boldmath $F$}\left(\mbox{\boldmath $X$}_i\right)
  +\mbox{\boldmath $K$} \left(\overline{\mbox{\boldmath $X$}}
  \left(t\right)
			     -\mbox{\boldmath $X$}_i\left(t\right)\right)
			     +\mbox{\boldmath $G$}\left(t\right)
			     +f \mbox{\boldmath $\eta$}_i\left(t\right),
 \label{eq.Gmodel}
\end{equation}
where the individual dynamics $\dot{\bd{X}}=\bd{F}(\bd{X})$ is assumed 
to be non-chaotic, and $\mbox{\boldmath $G$}\left(t\right)$ represents 
the coherent random force independent of $i$. By virtue of the $\bd{G}(t)$ 
term, the local Lyapunov exponent now fluctuates randomly, so that the 
anomalous fluctuation in $r$ could be recovered. This can be demonstrated 
with a population of oscillatory/excitable units of the FitzHugh-Nagumo 
type\cite{fitzhugh61,nagumo62}. The specific form of the model is given by
\begin{eqnarray}
\frac{dX_i}{dt}&=&\frac{1}{\epsilon}\left(X_i-X_i^3-Y_i\right)
    +K\cdot \left(\overline{X}-X_i\right) \nonumber\\
    &+&G_0\sin\phi(t)+f\cdot \eta_{i,X}(t),
    \nonumber\\
   \frac{dY_i}{dt}&=&aX_i+b+K\cdot \left(\overline{Y}-Y_i\right)
   +f\cdot \eta_{i,Y}(t).
\label{eq.FN}
\end{eqnarray}
where $G_0\sin\phi(t)$ represents external random force with $\phi$ generated 
from the dynamics of a random walker:
\begin{equation}
 \frac{d^{2}\phi}{dt^2}=-\gamma\frac{d\phi}{dt}+g\cdot\eta~.
  \label{eq.ranwalk}
\end{equation}   
Some numerical results of the above model are shown in
Figs. \ref{fig.fn.size} and \ref{fig.fn.pow} where 
the parameter values are so chosen that the individual unit is 
non-oscillatory but excitable. Nontrivial power-law dependence of 
$\langle r\rangle$ on $f$ with parameter-dependent exponent is again 
confirmed. Similar behavior of $\langle r\rangle$ of course persists when 
the individual dynamics becomes oscillatory.

\section{Summary}\label{sec.summ}

In the present paper, we argued that in globally coupled systems of 
nonlinear dynamical units small noise or imposed heterogeneities can 
generally cause anomalously strong dispersion of the synchronized clusters 
in the phase space. This was demonstrated numerically with a number of 
population models with chaotic dynamical units. The numerical results 
were explained theoretically in terms of a multiplicative stochastic 
process with additive noise. It turned out that the crucial condition 
for the occurrence of such anomaly is the random fluctuations of the local 
Lyapunov exponent associated with the individual units. This fact suggested 
some possible generalizations of the class of systems capable of exhibiting 
similar behavior. In particular, chaotic nature of the individual dynamics 
seemed unnecessary provided the population is subjected to a common random 
drive, and this was actually demonstrated with the population of the 
FitzHugh-Nagumo type excitable units. 



\end{document}